# Evidence for a Bulk Complex Order-Parameter in $Y_{0.9}Ca_{0.1}Ba_2Cu_3O_{7-\delta}$ Thin Films

E. Farber,[1,2] G. Deutscher,[1] B. Gorshunov,[3,4] and M. Dressel[3]

[1] School of Physics and Astronomy, Raymond and Beverly Sackler Faculty of Exact Science, Tel Aviv University Ramat Aviv 69978, Israel

[2] Department of Electrical and Electronics Engineering College Of Judea and Samaria, Ariel, Israel.

[3] 1. Physikalisches Institut, Universität Stuttgart, Pfaffenwaldring 57, D-70550 Stuttgart, Germany

[4] General Physics Institute, Russian Academy of Sciences, Moscow, Russia

We have measured the penetration depth of overdoped $Y_{0.9}Ca_{0.1}Ba_2Cu_3O_{7-\delta}$ (Ca-YBCO) thin films using two different methods. The change of the penetration depth as a function of temperature has been measured using the parallel plate resonator (PPR), while its absolute value was obtained from a quasi-optical transmission measurements. Both sets of measurements are compatible with an order parameter of the form: $\Delta_{dx^2-y^2}+i\delta_{dxy}$, with $\Delta=14.5 \pm 1.5$ meV and $\delta=1.8$ meV, indicating a finite gap at low temperature. Below 15 K the drop of the scattering rate of uncondensed carriers becomes steeper in contrast to a flattening observed for optimally doped YBCO films. This decrease supports our results on the penetration depth temperature dependence. The findings are in agreement with tunneling measurements on similar Ca-YBCO thin films.

Recent tunneling reports indicate the existence of spontaneous surface currents in YBCO high $T_c$ superconductors [1]. These currents shift the energy of the low lying surface states [2,3], characteristic of a d-wave order parameter (OP); the zero bias conductance peak (ZBCP) splits due to these states. An important question that remains to be solved is whether the additional, imaginary component of the OP associated to these surface currents [4] occurs only at the surface or whether it is in fact a bulk property [5]. While a surface-associated imaginary component would still be compatible with a pure $d_{x^2-y^2}$-wave bulk OP, a bulk-associated imaginary component would not be. This distinction has important implications for our understanding of the high $T_c$ mechanism, since some of the proposed models require a pure bulk d-wave symmetry [6,7], while others do not [8,9,10].

A relatively small imaginary component in the bulk OP cannot be easily detected by the phase sensitive experiments [11] that have established the dominant $d_{x^2-y^2}$-wave character of the OP. This is because it would produce only a small change in the spontaneous half quantum flux characteristic of π junctions, which might fall well within the margins of errors of the measurement (particularly if the imaginary component has an $id_{xy}$ symmetry). Likewise, it would not be easily detected by low temperature heat capacity measurements whose behavior is often dominated by impurity effects.

The low temperature behavior of the penetration depth $\lambda(T)$, which was studied in great detail in optimally doped samples [12], shows that in that case the imaginary component is zero or very small [5]. Then $\lambda(T)$ can be well fitted by expressions form of the $d_{x^2-y^2}$-wave theory in the unitary limit. In thin films and impure samples the behavior of $\lambda(T)$ follows the law

$$\lambda(T) \propto \frac{\lambda(0)\ln(2)}{\Delta_0}\frac{T^2}{T+T^*} \quad (1)$$

which describes a crossover between a $T^2$ low temperature behavior, dominated by impurities, and a linear behavior for $T > T^* \cong 0.83(\Gamma\Delta_0)^{1/2}$ in the unitary limit, where $\Gamma$ is the scattering rate and $\Delta_0$ is the zero temperature energy gap.

In the present research, overdoped Ca-YBCO thin films were measured in order to investigate the possible existence of a bulk complex order parameter. Our data for 10% Ca doped films having a $T_c$ of 85 K cannot be fitted by Eq. 1. On the contrary, they can well be fitted by a modified $d_{x^2-y^2}$-wave behavior, taking into account the non-zero imaginary component ($is$ or $id_{xy}$) of 1.8 meV.

Overdoped c-axis Ca-YBCO thin films were grown on (100) oriented $LaAlO_3$ substrates. The films were deposited by off-axis dc-sputtering, using a pressure template process [13]. For PPR measurements we have used films with a thickness of 3000 Å. X-ray diffraction reveal no secondary phases, except for a small broadening of the $LaAlO_3$ peak which suggest a (100) contribution from a-axis grains on the surface. The contribution of this broadening to the LaAlO$_3$ peak is about 1% of this peak.

The measurements were performed on two main sets of Ca-YBCO thin films, which were grown at two different temperatures. The higher temperature grown films exhibit a very small amount of a-axis grains, appearing at the surface as seen by the Scanning Electron Microscope (SEM) picture (see inset of fig. 2). The lower temperature grown films show a higher density of a-axis grains on the surface, and therefore can be regarded as more disordered. Atomic Force Microscope (AFM) pictures confirm the X-ray data and the SEM picture. They show a small density of a-axis grains at the surface, which causes an average roughness of about 150 Å in the higher temperature grown films, while in the lower temperature grown set, it causes a surface roughness of 350 Å. A comprehensive sample characterization and experimental details are given elsewhere [14].

The transition temperature was measured using an inductive method. The films with a smaller amount of a-axis grains and smaller roughness exhibit $T_c$=84.5 K and a transition width smaller than 0.5 K. The other set of films exhibits $T_c$=74 K and a transition width of about 1.5 K, which can be associated with the presumed a-axis grains on the surface.

For the measurements of $\lambda(0)$ we have used a much thinner overdoped Ca-YBCO thin films with a thickness of 350 Å, $T_c$ = 61 K and a transition width of 1.5 K. These lower $T_c$ thin films can be regarded as slightly disordered and therefore are giving an upper bound value of $\lambda(0)$ for overdoped thin films [17]. We have used the Mach-Zehnder arrangement in which the absolute value of the transmission coefficient and the phase change of the transmitted wave are measured for a film on the substrate, from which the complex dielectric function of the film and the absolute value of the penetration depth can be obtained [15]. Fig. 1 shows the so obtained frequency dependence of the dielectric constant of the film in the superconducting state, at 2 K. It displays a dielectric response typical for a superconductor which is connected with the delta-function at zero frequency in the conductivity spectrum [16]; it can be fitted with $\varepsilon_1 = -\omega^2_p/\omega^2$ with $\omega_p$ being the plasma frequency of the superconducting condensate. The best fit (line in Fig.1) allows to determine the plasma frequency of condensed carriers and the penetration depth: *$\omega_p$ = 48380 cm$^{-1}$, $\lambda(0) = 2070 \pm 200$ Å*. This value is consistent with the results reported by Bernhard et al. using TF-µSR measurements [17].

The change of the penetration depth *Δλ(T) (Δλ(T) = λ(T) - λ(T$_0$))* where $T_0$ is the onset measurement temperature) is determined by the PPR method. A typical Q value of the PPR in these measurements was about 13000. This yields a penetration depth change measurement resolution of about 0.1 Å. Fig. 2 shows the $\lambda(T)$ dependence assuming the value for $\lambda(0)$ determined by the quasi-optical method. These films are almost free of a-axis grains, a SEM picture is shown in the inset of the figure. Flattening of $\lambda(T)$ at low temperatures is clearly seen followed by a linear behavior at higher temperatures.

Fitting the penetration depth variation data, $\Delta\lambda(T)$, by using a $d_{x^2-y^2}$-wave gap within the unitary limit, eq. 1, yields a crossover temperature $T^* = 32$ K and a $d_{x^2-y^2}$-wave slope $d\lambda/dT$ of 31 Å/K. Assuming a 14.5 meV gap, as expected for $T_c = 84.5$ K [18], this slope results in a large zero temperature penetration depth for this system, $\lambda(0) = 3250 \pm 250$ Å taking thickness corrections into account, $f(t/\lambda) = coth(t/\lambda) + t/\lambda(sinh^2(t/\lambda))^{-1}$ where $t$ is the film thickness [14]. That value is much larger than the value obtained directly by the quasi-optical method. These results are in contrast to those we obtained on optimally doped YBCO thin films, which show a $T_c = 92$ K. Fitting these data by a $d_{x^2-y^2}$-wave OP within the unitary limit yields a crossover temperature $T^* = 16$ K and a slope of 3 Å/K [19] which is in agreement with the predictions of $d_{x^2-y^2}$-wave theory [20].

As the next step in the analysis, $\Delta\lambda(T)$ was fitted using a complex order parameter. The fit was carried out using a numerical calculation for the penetration depth variation due to thermal excitations, relying on the following BCS equation:

$$\left[\frac{\lambda(0)}{\lambda(T)}\right]^2 = 1 - \frac{2}{k_B T} \int_0^\infty dE \frac{N(E)}{N_0} f(E)[1-f(E)] \qquad (2)$$

where the normalized quasiparticles density of states $N(E)$ is related to the order parameter by the following relation:

$$\frac{N(E)}{N_0} = Re\left[\frac{E}{\left(E^2 - \Delta_{complex}^2\right)^{1/2}}\right] \qquad (3)$$

taking $\Delta_{complex} \propto .d_{x^2-y^2} + id_{xy}$, or $d_{x^2-y^2} + is$.

The best fit to the experimental data of $\lambda(T)$, shown in Fig. 2, is attained by using a density of states, which contains a complex order parameter of the form $\Delta_{dx^2-y^2} + i\delta_{dxy}$ where $\Delta=14.5$ meV and $\delta=1.8$ meV (an order parameter of the form $\Delta_{dx^2-y^2} + i\delta s$ can be suitable as well). We emphasize that these two values are obtained by fitting the data to Eq. 2, with $N(E)$ given by Eq. 3 and taking $\lambda(0) = 2070$ Å as obtained by the quasi-optical method. The low temperature flattening of $\lambda(T)$ for these overdoped films is in contrast to the power law of $\Delta\lambda(T)$ reported for optimally doped thin films [21,19].

Fig. 3 shows the surface resistance as a function of temperature for the set of samples with a negligible amount of a-axis grains on their surface. It presents a steeper change below 25 K as compared to the surface resistance of a set of pure YBCO films. This behavior is reminiscent of that reported by Klein et al. [22] on oxygen overdoped YBCO films and gives further support for the existence of a small subgap.

Fig. 4 shows the quasi-particles scattering rate as a function of temperature for the set of samples with a negligible amount of a-axis grains on their surface. It presents a steeper change of the scattering rate of uncondensed carriers below 15 K, which may be an additional indication for the opening up of a finite gap, i.e., the existence of a complex order parameter. The lower quality set of films shows a similar behavior for the scattering rate. These scattering rates temperature dependence contrasts with the almost constant value, observed in YBCO thin films below 30K, using the same analysis [23].

For the films which contain a high density of a-axis grains, $T_c$=74 K, we have not been able to obtain a satisfactory fit to the data neither in the $d_{x^2-y^2}$-wave unitary limit, nor using a complex order parameter (eq. 2 and eq. 3). There is some indication for a finite gap at low temperatures, where the experimental behavior of $\Delta\lambda(T)$ is better fitted to BCS exponential dependence than to a $T^2$ law. Using the BCS exponential dependence we obtained a finite gap of 1.7 meV, in agreement with the imaginary component obtained by using a complex order parameter. It is also possible to fit the low temperature behavior to a $T^3$ law as reported for optimally doped PrCeCuO [24]. At higher temperatures there is a tendency to recover a linear temperature dependence. One can regard the a-axis grains as a sort of imperfection, which exists on the c-axis thin films surface. This suggests that in the presence of a substantial concentration of imperfections, it may be necessary to use a theory involving a complex order parameter in the unitary limit to analyze the data.

In conclusion, our measurements on overdoped Ca-YBCO films show for the first time that in this compound $\lambda(T)$ is best fitted by a complex order parameter of the form $\Delta_{dx^2-y^2}+i\delta_{dxy}$. This is in agreement with tunneling results on overdoped thin films [25,26], giving a firm indication for a bulk complex order parameter with a small imaginary component $id_{xy}$ or $is$ in overdoped YBCO thin films. An attempt to fit $\lambda(T)$ using a $d_{x^2-y^2}$- wave order parameter within the unitary limit yields a slope of 31 Å/K which gives $\lambda(0) = 3250 \pm 250$ Å, much larger than the value measured directly by the quasi–optical method. This behavior is in contrast with that of optimally doped YBCO films, for which a pure $d_{x^2-y^2}$ order parameter fits well the data assuming unitary limit scattering. A further support for a finite gap can be seen by the steeper reduction of the scattering rate below 15 K, which contrasts with the almost constant value for optimally doped thin films.

## ACKNOWLEDGMENTS

The authors (E. Farber and G. Deutscher) would like to thank Nicole Bontemps for the fruitful discussion. This work was supported in part by the Heinrich Hertz-Minerva Center for High Temperature Superconductivity, and by Oren Family chair of Experimental Solid State Physics.

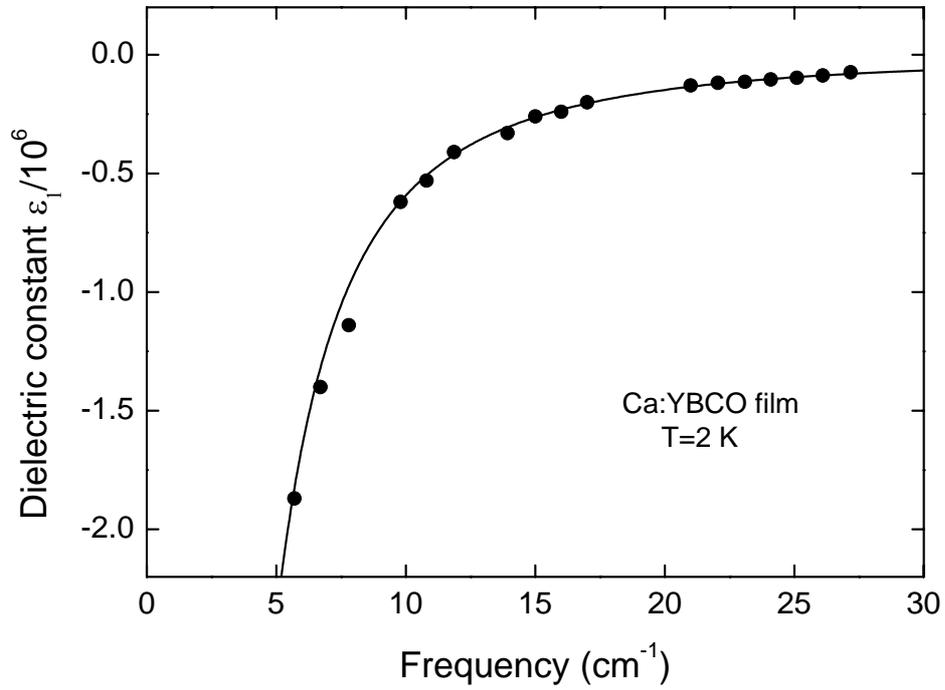

Fig. 1. The real part of the dielectric function versus frequency for a thin Ca overdoped YBCO film, T= 2 K. Dots show experiment, lines show the fit to expression $\varepsilon_1 = -\omega_p^2/\omega^2$ describing the response of superconducting condensate [16], as discussed in the text. The best fit yields *λ(0) = 2070 Å*.

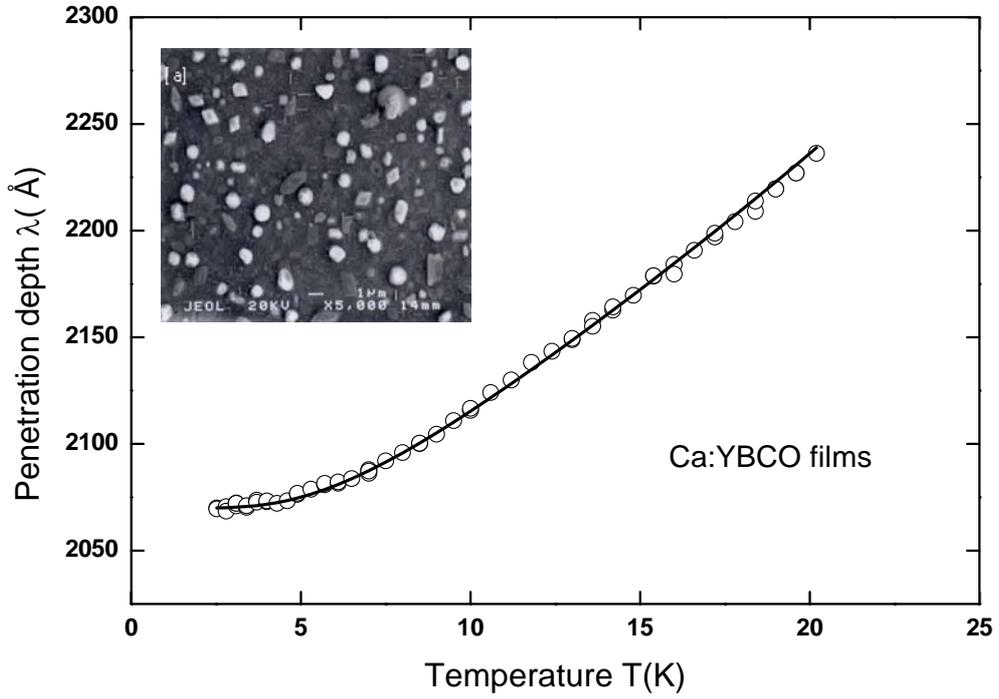

Fig.2. The penetration depth versus temperature for a Ca overdoped set of YBCO samples having $T_c = 84.5\ K$. Open circles show experimental data and the solid line represents a fit using equation 2 and 3 with a complex order parameter of the form $\Delta_{dx^2-y^2} + i\delta_{dxy}$ where $\Delta = 14.5\ meV$ and $\delta = 1.8\ meV$. That fit yields $\lambda(0) = 2070\ Å$. The inset is a typical SEM picture of these films showing a negligible density of a-axis grains on their surface.

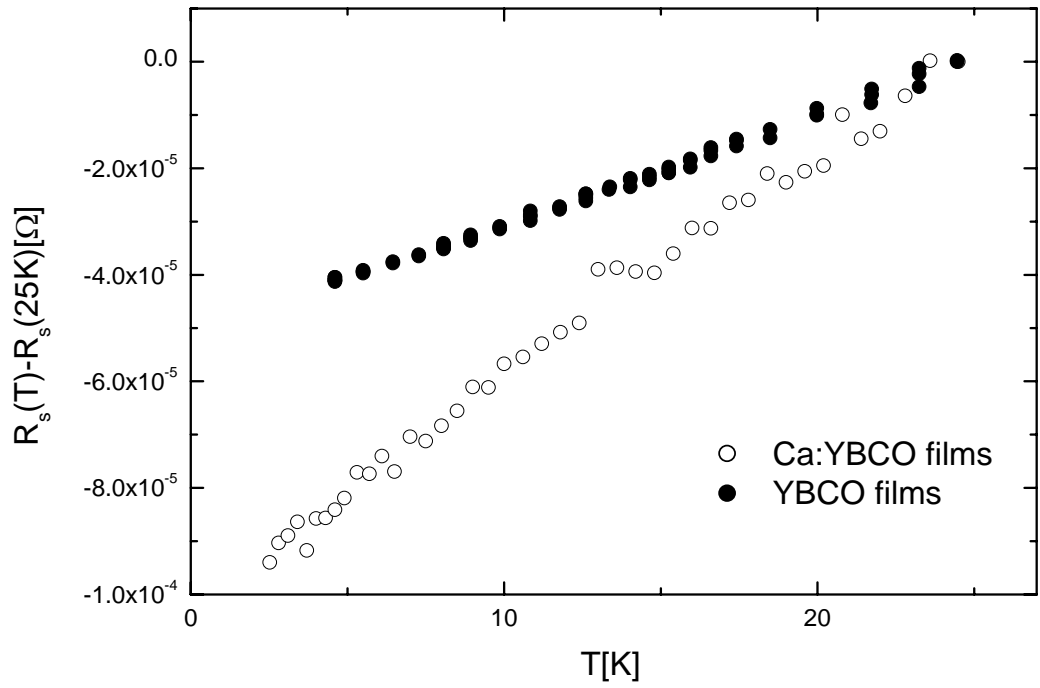

Fig. 3. The change of the surface resistance for two types of samples: open circles for overdoped Ca-YBCO films, filled circles for optimally doped YBCO films.

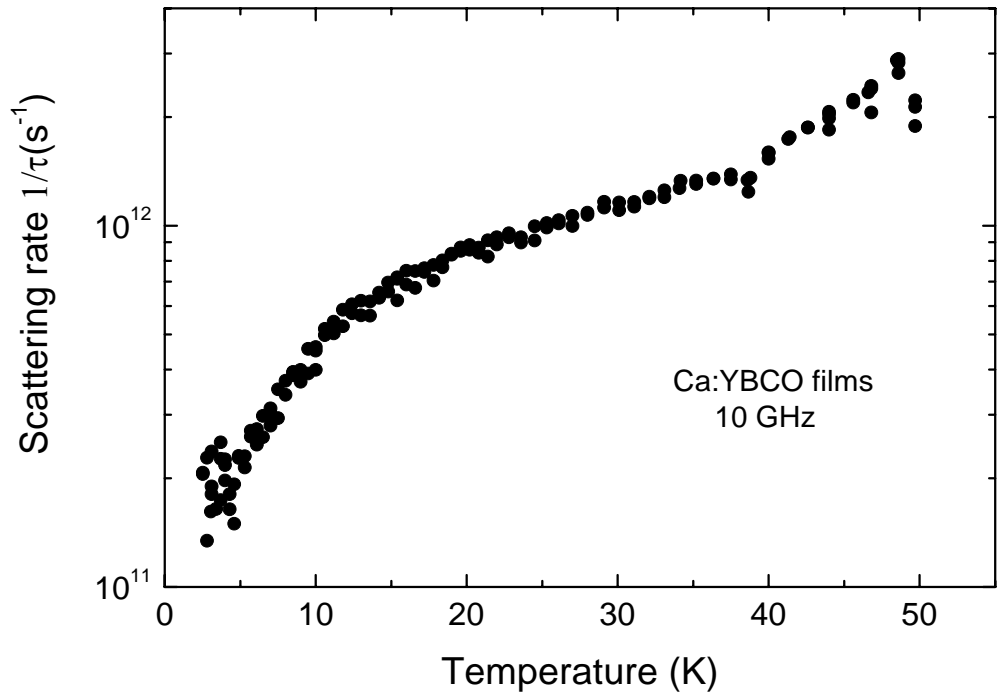

Fig. 4. Temperature dependence of the scattering rate of uncondensed carriers for the films which are almost free of a-axis grains. The data were taken at *10 GHz*.